\documentclass[pre,twocolumn]{revtex4-1}

\usepackage{amsmath}
\usepackage{enumerate}
\usepackage{graphicx}
\usepackage{mathbbol}
\usepackage{amsfonts}
\usepackage{natbib}

\begin{document}


\title{L\'evy flights on the half line} 



\author{Reinaldo Garc\'ia-Garc\'ia}
\email[]{reinaldo.garcia@cab.cnea.gov.ar}
\affiliation{Centro At\'omico Bariloche, 8400 S. C. de Bariloche, Argentina}

\author{Alberto Rosso}
\email[]{alberto.rosso@u-psud.fr}
\affiliation{Laboratoire de Physique Th\'eorique et Mod\`eles Statistiques, Universit\'e Paris Sud 11 and CNRS}

\author{Gr\'egory Schehr}
\email[]{gregory.schehr@th.u-psud.fr}
\affiliation{Laboratoire de Physique Th\'eorique d'Orsay, Universit\'e Paris Sud 11 and CNRS}


\date{\today}

\begin{abstract}
We study the probability distribution function (pdf) of the position of a L\'evy flight of index $0<\alpha<2$ in  presence of an absorbing wall at the origin. The solution of the associated fractional Fokker-Planck equation can be constructed  using a perturbation scheme around the Brownian solution (corresponding to $\alpha = 2$), as an expansion in $\epsilon = 2 - \alpha$.  We obtain an explicit  analytical solution,  exact at the first order in $\epsilon$, which allows us to conjecture the precise asymptotic behavior of this pdf, including the first subleading corrections, for any $\alpha$. Careful numerical simulations, as well as an exact computation for $\alpha = 1$, confirm our conjecture.   
\end{abstract}

\pacs{}

\maketitle 

\section{Introduction}

Random walks are trajectories consisting of a collection of random steps. They are employed to model the stochastic activity observed in many fields such as physics, biology, quantitative finance or computer science. As such they have been widely studied by mathematicians \cite{SA54, feller, hughes_book} and physicists \cite{montroll_weiss, bouchaud_georges, klafter_review, satya_leuven}.  One of the simplest example is given by a symmetric one dimensional random walker whose  position, $x(n)$, after $n$  steps evolves, for $n \geq 1$, according to  
\begin{eqnarray}\label{def_rw}
x{(n)} = x{(n-1)} + \eta(n) \;,
\end{eqnarray} 
starting from $x(0) = 0$. Moreover we  consider random steps  {\em  independent} and {\em identically}  distributed, according to  a  probability distribution  $\varphi(\eta)$: consequently the random walk is {\em Markovian} and {\em homogeneous}.

 Despite being simple to define, most of the properties of a random walk remain difficult to determine analytically. However,  when the number of steps is large, the random walk displays a "universal" behavior and the statistics of the position $x(n)$ becomes independent of most of the details of $\varphi(\eta)$.  This asymptotic regime is the one we have more chance to characterize and it is often the one which is relevant for  applications. In this limit, two cases should be distinguished.  If $\varphi(\eta)$ is a narrow distribution the random walk exhibits a diffusive behavior, $x(n) \sim  n^{1/2}$ and only  the second moment  of the step distribution, $\int_{-\infty}^\infty \eta^2 \varphi(\eta) d\eta = \sigma^2 $,  affects the statistics of $x(n)$. On the other hand  if the random variables $\eta(n)$'s have a broad distribution, with a diverging second moment, {\it i.e.} 
\begin{eqnarray}\label{jump_tail}
\varphi(\eta) \sim \frac{c}{|\eta|^{1+\alpha}} \;, \; |\eta| \gg 1 \;, 
\end{eqnarray}
with $0 < \alpha < 2$, the random walk exhibits a super-diffusive behavior $x(n) \sim n^{1/\alpha}$.  Such power law distributions (\ref{jump_tail}) have been initially studied in the early sixties in economy \cite{pareto} and finance \cite{mandelbrot} and in the early eighties they started to proliferate in physics where they have found many applications ranging from disordered and glassy systems, super-diffusion in micellar systems, laser-cooling of cold atoms \cite{shlesinger_book}, random matrices \cite{biroli}, photons in hot atomic vapours~\cite{mercadier} etc... One striking feature of such processes  is that their statistical behavior is governed by a few rare events, whose occurrence are thus governed by the {\it tail} of the distribution.  

Also in this case, when the number of steps becomes large,  we expect that the statistics of $x(n)$  becomes independent of the details of $\varphi (\eta)$ except for the index $\alpha$ and the constant $c$. In particular, in absence of boundaries, the central limit theorem ensures that  the propagator, $P_{}(x,n)$, {\it i.e.} the probability to find the particle in $x$ after $n$ steps, converges to a stable distribution  given by:
\begin{eqnarray}\label{def1}
&&P_{}(x,n) = \frac{1}{n^{1/\alpha}}R\left( y\right) \;, \\
&& \int_{-\infty}^\infty R(y) e^{i k y} \, dy = e^{-|a k|^\alpha} \;, \label{char_func}
\end{eqnarray}
where the  parameter $a$ is related to the constant $c$ [see Eq.~(\ref{R_intro_2}) below] and  where we have introduced, from dimensional analysis,  the rescaled variable 
\begin{eqnarray}\label{def_rescaled}
y \equiv \frac{x}{n^{1/\alpha}} \;.
\end{eqnarray}
 Although the Fourier transform of $R(y)$, its characteristic function, has a very simple expression (\ref{char_func}), there is no
simple closed form expression for $R(y)$, except for $\alpha = 1$ which corresponds to Cauchy distribution. It however admits the following asymptotic expansion, valid for any value of $\alpha$~\cite{hughes_book}: 
\begin{eqnarray}\label{free_levy_intro}
R(y) \sim \frac{1}{\pi |y|} \sum_{k=1}^\infty a^{\alpha k}\frac{\sin{\left(\frac{\alpha k \pi}{2} \right)} \Gamma(\alpha k +1)(-1)^{k+1} }{k! |y|^{\alpha k}} \;.
\end{eqnarray}
One thus sees on this expression that $R(y)$ inherits the power law tail of the step distribution (\ref{jump_tail}). One can further show \cite{hughes_book, feller} that the amplitude itself is not renormalized such that to leading order   
\begin{eqnarray}\label{R_intro}
R(y) \sim \frac{c}{y^{1+\alpha}} + \frac{d}{y^{1+2 \alpha}} + {\cal O}(y^{-{1-3\alpha}}) \;,
\end{eqnarray}
which fixes the value of $a$
\begin{eqnarray}\label{R_intro_2}
 a^\alpha= \frac{\pi \, c}{  \sin \left(\frac{\alpha \pi}{2} \right) \Gamma(\alpha +1)} \;,
\end{eqnarray}
while $d$ can also be obtained explicitly from Eq. (\ref{free_levy_intro}).

Although free L\'evy flights are thus perfectly well understood, there are physical situations which actually involve L\'evy flights in a confined geometry. An interesting example is the L\'evy flight model which has been proposed \cite{davis_marshak} to describe the transports of solar photons in cloudy atmosphere. These photons are eventually reflected back to space or absorbed by the ground, so their trajectories are bounded random walks. In such cloudy atmosphere, whose heigth is typically of the order of $10$ kms, the photons can be trapped in optically dense region (inside the clouds), travelling less than a meter between scatterings, while they can "fly" many kilometers from cloud to cloud. It was shown experimentally that L\'evy flights provide a reliable description of the photons transport in such situations~\cite{pfeilsticker}. It was shown, in addition, that the one-dimensional model is a reasonable approximation of the three-dimensional geometry~\cite{davis_pfeilsticker}. More recently, L\'evy flights in confined geometry have also found applications in the context of random search problems~\cite{jpa_stanley}.

Obviously, when the walker is confined inside a domain, the central limit theorem does not apply, but the scaling analysis is still valid and the "universal" behavior of the rescaled position $y=x/n^{1/\alpha}$ is still expected. Computing the statistics of the rescaled variable $y$ in presence of confinement is in general possible for Brownian motion ($\alpha=2$) for which powerful analytical tools are available such as path-integral techniques \cite{yor_functional, satya_functional}. Unfortunately, for  L\'evy flights, analytical approaches are usually quite difficult.  Recently a lot of papers advertised the  possibility to write a Fractional Fokker-Planck  equation for  L\'evy flight propagator  \cite{klafter_review} :
\begin{eqnarray} \label{fp_intro}
 \frac{\partial}{\partial t} P(x,t) = a^\alpha \frac{\partial^{\alpha}}{\partial |x|^\alpha} P(x,t) \;, \; P(x,t=0) = \delta(x) \;,
\end{eqnarray}
where the continuum time $t$ captures the large $n$ behavior of the random walk and the fractional operator, $\frac{\partial^{\alpha}}{\partial |x|^\alpha}$, is the Riesz-Feller derivative of fractional order $\alpha > 0$ \cite{IP99, SKM93}, which has an integral representation involving a singular kernel of power-law form. In absence of boundaries this equation can be simply written in Fourier space
\begin{eqnarray} \label{fp_k}
 \frac{\partial}{\partial t} \tilde P(k,t) = - |a\, k|^\alpha \tilde P(k,t) \;, \:  \tilde P(k,t=0) = 1 \;,
\end{eqnarray}
and  it is easy to check that the free propagator introduced in Eq.~(\ref{def1}) is a solution of Eq.~(\ref{fp_k}), (with the identification, $t \to n \gg 1$).

In presence of boundaries  the translational invariance is broken and the Fourier representation becomes useless. For  $\alpha=2$, when the fractional operator becomes the standard Laplacian, the method of images  allows to express the propagator in presence of boundaries as a linear combination of free propagators. Unfortunately we will see that these techniques  can not be applied, for $\alpha<2$~\cite{ZK95, metzler_images}.  More generally, if  translational invariance is lost, the fractional Fokker-Planck equation in Eq.~(\ref{fp_intro}), becomes a difficult integro-differential equation with non-local boundary conditions. One could conclude that the Fokker-Planck formalism is of little help for L\'evy flights. However, following a recent work by Zoia, Rosso and Kardar \cite{ZRK07}, we show that  Eq. (\ref{fp_intro} ) can be studied also in presence of boundaries using a  perturbation theory where the small parameter $\epsilon \ll 1$ is $\alpha = 2 - \epsilon$. At variance with the method of Ref. \cite{ZRK07} we perform this perturbation theory directly in the continuum limit without resorting to a discretization (in space) of the trajectories. The calculations turn out to be somewhat simpler in this continuum setting. 


\begin{figure}[ht]
\includegraphics[width = \linewidth]{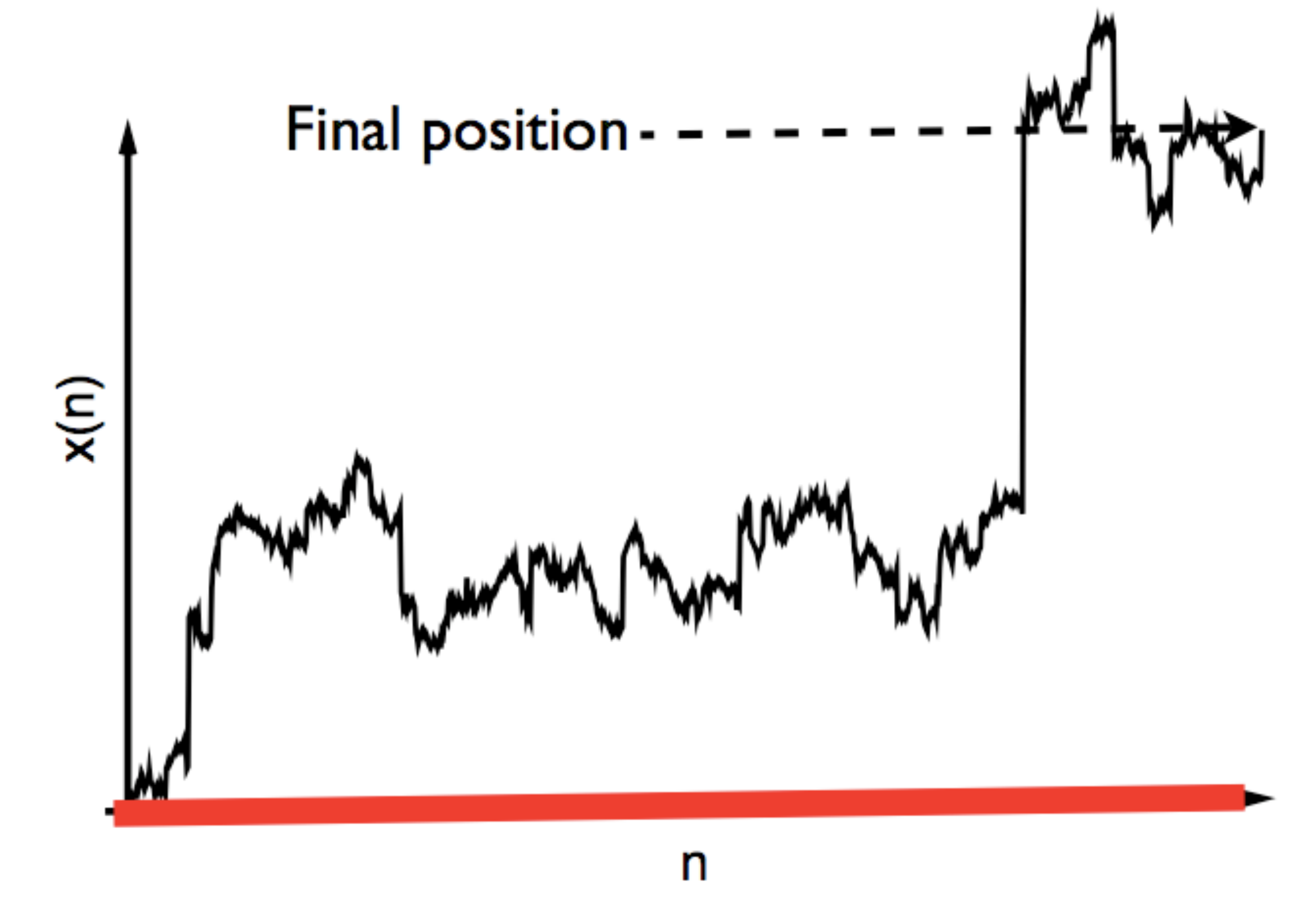}
\caption{Trajectory of a random walker with an absorbing boundary condition at the origin.  $P_+(x,t)$, computed perturbatively in Eq.~(\ref{W_elementary}), is the probability density function of the position of such a random walker at time $t$.}\label{fig_absorbing}
\end{figure}

For concreteness we will study in detail  the case where there is an absorbing wall in $x=0$ as depicted in Fig. \ref{fig_absorbing}: we thus consider only the paths that remain positive up to the $n^{\rm th}$ step. In the limit of large $n$ the probability density function to find the particle in $x$ after $n$ time steps also takes the scaling form, as in Eq. (\ref{def1}):
\begin{eqnarray}
P_+(x,n) = \frac{1}{n^{1/\alpha}}R_+\left( y\right) \;.
\end{eqnarray}
We emphasize that $R_+(y)$, being a probability density function, is normalized, {\it i.e.} $\int_0^\infty R_+(y) \, dy = 1$, while, in Ref. \cite{ZK95}, Zumofen and Klafter studied a similar quantity, which is however not normalized (see also Ref.~\cite{ivanov}). In particular, they were able to show that the small argument behavior of $R_+(y)$ is given by \cite{ZK95, zoia_semi_inf}
\begin{eqnarray}
R_+(y) \sim y^{\alpha/2} \;,
\end{eqnarray}
in contrast with the method of images that would predict $R_+(y) \sim y$. Our perturbative approach allows us to conjecture the exact behavior of the tail of $R_+(y)$, which controls the statistics of rare events~: 
\begin{eqnarray}\label{result_large}
R_+(y) = \frac{c_+}{y^{1+\alpha}} \; + \; 
\begin{cases}
&\frac{d_+}{y^{2+\alpha}} + o (y^{-2-\alpha})\;, \; 2 > \alpha > 1 \;, \\
& \\
&\frac{d_+}{y^{1+2\alpha}} + o(y^{-1-2 \alpha}) \;,\; 1 > \alpha > 0 \;.
\end{cases}
\end{eqnarray}  
To our knowledge, only the exponent of the leading term $R_+(y) \propto y^{-1-\alpha}$, was known from Ref. \cite{ZK95}. Here we obtain the  
exact result for the amplitude $c_+$ 
 \begin{eqnarray}\label{relation_c}
 c_+ = 2 c \;,
 \end{eqnarray}
 where $c$ is the amplitude of the tail of the jump distribution (\ref{jump_tail}). In addition, the first sub-leading corrections in Eq. (\ref{result_large}), by comparison to the free case (\ref{R_intro}), also bear the fingerprints of the absorbing wall. These results (\ref{result_large}) are first obtained analytically for $\alpha$ close to $2$, {\it i.e.} $\alpha = 2 - \epsilon$, using a perturbation theory to first order in $\epsilon$. We also obtain this behavior for $\alpha = 1$, albeit with logarithmic corrections for the subleading term, for which an exact calculation can be done. We then demonstrate this behavior using thorough numerical simulations.

The paper is organized as follows. In Section II we present the general framework of the perturbation scheme. We first illustrate it on the simplest example of the free propagator in Section II A, and then in Section II B we study the case with an absorbing boundary at the origin. The discussion of the results is left in Section II C. Section III contains the results of our numerical simulations and our conclusions are in section IV. The Appendices A, B, C contain some technical details.

\section{Perturbation scheme}

In this section, we set $a=1$ for simplicity and without loss of generality, and write  
the Fractional Fokker-Planck equation (\ref{fp_intro}) in the familiar Schr\"odinger form
\begin{eqnarray}\label{def_H}
&&\partial_t P(x,x_0,t) = {\cal H} P(x,x_0,t) \;, \\
&& P(x,x_0,t=0) =\delta(x-x_0) \;,
\end{eqnarray} 
where the propagator $P(x,x_0,t=0)$ represents the probability density to find a particle in the interval $[x, x+ d x]$ at time $t$, knowing that the particle was in $x_0$ at time $0$ and the operator ${\cal H} $ is the fractional operator of index $\alpha$. In Quantum Mechanics Eq.~(\ref{def_H}) corresponds to the Schr\"odinger equation of the element $(x,x_0)$ of the density matrix $P$ at the temperature $1/t$. The general  solution of Eq.\,(\ref{def_H}) reads
\begin{equation}\label{sol1}
P(x,x_0,t)=\int dq \, \psi_q^*(x_0) \psi_q(x) e^{E(q) t} \;,
\end{equation}
where $E(q)$ are the eigenvalues and $\psi_q(x)$ the associated eigenfunctions of the operator ${\cal H}$. They are  solutions of the eigenvalue problem, ${\cal H}  \psi_q(x) =E(q) \psi_q(x)$, {\em with the appropriate boundary conditions} and
satisfy  the orthonormality and closure relations:
\begin{eqnarray}
&&  \int  \psi_q(x) \psi_{q'}^*(x) dx = \delta(q-q') \;, \label{rel1} \\
&& \int  \psi_q(x) \psi_{q}^*(x') dq =  \delta(x-x') \label{rel2} \;.
\end{eqnarray}
Please note that the domain of integration over $x$ depend on the boundary condition and the integration over $q$ is meant over the all spectrum. If the spectrum is discrete the integral is replaced by a discrete sum.

The eigenvalue problem of the operator ${\cal H} $ is in general very difficult and the solution is known only in the absence of boundaries where the  eigenfunctions are simple plane waves. The case $\alpha=2$ corresponds to the standard Laplacian for which a lot of results are known. In particular the spectrum can be exactly solved in one dimension. A possible strategy  to gain insights on the behavior of a L\'evy flights in presence of boundaries is to use the perturbation scheme around the well studied $\alpha=2$ case. For $\alpha=2-\epsilon$ the operator ${\cal H} $ and the propagator can be expanded in powers of $\epsilon$,
\begin{eqnarray}
&&P(x,x_0,t)=P^{(0)}(x,x_0,t)-\epsilon P^{(1)}(x,x_0,t)+{\cal O}(\epsilon^2)   \label{sol2} \; \nonumber \\
&& {\cal H}=  {\cal H}_0-\epsilon  {\cal H}_1+O(\epsilon^2) \label{def_H2} \;,
\end{eqnarray} 
where $P^{(0)}(x,x_0,t)$ is the propagator associated to ${\cal H}_0=  \partial_x^2$, with the prescribed boundary conditions.
The expression of the first correction $P^{(1)}(x,x_0,t)$ in Eq. (\ref{sol2}) is well known from Quantum Mechanics and it is re-derived in Appendix \ref{a_1}. It reads, writing $P^{(1)} \equiv P^{(1)}(x,x_0,t)$:
\begin{equation}\label{sol2_bis}
P^{(1)}= \int_{q} \int_{q'}  \psi_q(x) \psi_{q'}^*(x_0) \frac{e^{E(q) t}-e^{E(q') t}}{E(q')-E(q)}     \langle  q |{\cal H}_1| q'   \rangle , 
\end{equation}
where we use the notation $\int_q \equiv \int dq$ and where it is understood here that $\psi_q(x)$ are the eigenvectors and $E(q)$ the corresponding eigenvalues of ${\cal H}_0$ and for  the matrix elements $\langle  q |{\cal H}_1| q' \rangle$ we use the bra-ket notation, borrowed from Quantum Mechanics with $\langle x | q \rangle = \psi_q(x)$. This formula (\ref{sol2_bis}) is the cornerstone of the perturbation approach presented here. 

\subsection{Absence of boundaries}
We first illustrate this perturbative approach by computing the first order correction to the Brownian propagator in absence of boundaries, {\it i.~e.} the Gaussian propagator, when  $\alpha=2-\epsilon$ and $x_0=0$.  In this simple case the eigenvalue problem of ${\cal H}_0= \partial_x^2$ gives:
\begin{equation}
\psi_q(x)=\frac{1}{\sqrt{2 \pi}} e^{i q x}, \; E(q)=- q^2, \;  -\infty<q<\infty .
\end{equation}
Using that $ k^{2 - \epsilon} = k^2 - \epsilon \, k^2 \log |k| + {\cal O}(\epsilon^2) $, the matrix element $\langle  q |{\cal H}_1| q' \rangle$ can be explicitly computed:
\begin{eqnarray}\label{H1}
\langle q| 	{\cal H}_1| q' \rangle  = \iiint  d x_1 \, d x_2 \, \frac{d k}{2 \pi}  \psi_q^*(x_1)    \psi_{q'}(x_2)    \nonumber \\
 \times \,   k^2 \log|k|   e^{i k (x_1-x_2)}=\delta(q-q') q^2 \log|q|,
\end{eqnarray}
where the integral over $x_1, x_2, k$ is performed over the whole real axis. Note that, thanks to the oscillating term $e^{ik(x_1-x_2)}$, the integral
over $k$ in (\ref{H1}) is dominated by the small values of $k$ where one can safely expand $k^{2-\epsilon}$ in powers of $\epsilon$.  
From Eq. (\ref{sol2_bis}) one obtains: 
\begin{equation} \label{sol2b}
P_{}^{(1)}(x,t)=-  t \,\int_{0}^\infty \,  \frac{d q }{ \pi} \,  \cos{ q x} \,  q^2 \,  \log q \,  e^{-q^2  t}.
\end{equation} 
Using $k=q \sqrt{t}$ and the scaling variable $z=x/\sqrt{t}$, Eq.~(\ref{sol2b}) can be recast in a simpler form
\begin{eqnarray}
 &&\sqrt{t} \, P_{}^{(1)}(x,t)= R_A(z)+R_B(z) \log t   \;, \label{premiere} \\
&&R_A(z)=   - \int_{0}^\infty  \,  \frac{d k }{ \pi} \, \cos{k z}\, k^2 \,\log k \,e^{-k^2}  \;, \label{firstint} \\
&&R_B(z)=\int_{0}^\infty \,  \frac{d k }{2 \pi} \, \cos{k z}\, k^2\, e^{-k^2} \;. \label{secondint}
\end{eqnarray} 

From scaling argument one expects (\ref{def1})
\begin{eqnarray}\label{scaling_free}
P(x,t) = \frac{1}{t^{1/\alpha}} R\left(\frac{x}{t^{1/\alpha}} \right) \;,
\end{eqnarray}
and a general issue of these perturbative computations is that  for $\alpha=2$ the natural scaling variable is $z=x/\sqrt{t}$, while for $\alpha=2-\epsilon$ the correct scaling variable is $y=x/t^{\frac{1}{2-\epsilon}}$, which also admits a perturbative expansion. For this reason our
final result should be recast  in terms of $y$, in order to identify the perturbative expansion of the scaling function $R(y)$ in Eq. (\ref{scaling_free}) as
\begin{eqnarray}\label{free_exp}
R(y)=R^{(0)}(y)-\epsilon R^{(1)}(y) + {\cal O}(\epsilon^2) \;, 
\end{eqnarray}
where  $R^{(0)}(y)$ is the Gaussian propagator given by
\begin{eqnarray}
R^{(0)}(y) = \frac{1}{2 \sqrt{\pi}} e^{-y^2/4} \;.
\end{eqnarray}
This can be done if  in the equation  $ t^{1/\alpha} P(x,t)=   R(y)$ we expand  at the first order in $\epsilon$ both $t^{1/\alpha} \sim \sqrt{t}+ \frac{\epsilon}{4}  \sqrt{t} \log t$ and $y \sim z  +\frac{1}{4}\epsilon z \log t$. After some simple algebra we can write 
\begin{equation}
\sqrt{t} P_{}^{(1)}(x,t)= R^{(1)}(z) + \frac{\epsilon}{4} \log t \left(1+ z \partial_z\right) R^{(0)}(z) \;. 
\end{equation}
 Comparing  with Eq.~(\ref{premiere})  we identify $ 4 R_B(z)= R^{(0)}(z)+ z \partial_z R^{(0)}(z) $ so that we conclude that $R^{(1)}(y)$ in Eq. (\ref{free_exp}) is given by:
\begin{eqnarray}
 && R^{(1)}(y) =   - \int_{0}^\infty  \,  \frac{d k }{ \pi} \, \cos{k y}\, k^2 \,\log k \,e^{-k^2}\;.  \label{firstintb}
\end{eqnarray} 
By performing an asymptotic analysis of Eq.~(\ref{firstintb}) for large $y$ one finds a series expansion of $R(y)$ given in Eq. (\ref{free_exp}) which converges nowhere but exists as a formal power series:
\begin{equation} \label{free_asymp}
R(y) \sim  \epsilon  \sum_{k=1}^\infty \frac{(2 k)!}{2 (k-1)!} \frac{1}{y^{2 k+1}}  \sim \epsilon \left( \frac{1}{y^3}+  \frac{12}{y^5}+ \frac{180}{y^7}+\ldots \right). \nonumber
\end{equation}
This result in in perfect agreement with the expansion given in Eq. (\ref{free_levy_intro}) for $\alpha=2-\epsilon$.

\subsection{ Propagator in presence of an absorbing boundary at $x=0$}

We consider now $G_+(x,x_0,t) \, dx$, the probability to find a particle in the interval $[x, x+ d x]$ at time $t$, knowing that the particle was in $x_0$ at time $0$ and given that it stayed positive up to time $t$ (see Fig. \ref{fig_absorbing}). 
At variance with the free propagator, the integral over $x$ of  $G_+(x,x_0,t)$ is smaller than one and gives the fraction of surviving walker up to time $t$ ({\it i.e.} the survival probability). In the geometry defined by Eq. (\ref{def_rw}) the initial position of the discrete random walk is $x_0=0$. Here we are considering a process which is continuous in time and this initial condition $x_0 = 0$, together with the presence of an absorbing boundary at the origin is ill-defined. Indeed, it is well known that if the continuous time walker crosses zero once
it will re-cross zero infinitely many times immediately
after the first crossing. Therefore, it is impossible
to enforce the constraint $x_0= 0$ and simultaneously
forcing the position of the continuous time walker to be strictly positive immediately after. Therefore we set $x_0 >0$ and small in order to 
regularize the continuous time process and we will take the limit $x_0 \to 0$ at the end of the calculation. Our final result corresponds to the geometry of the Eq.~(\ref{def_rw}) in the limit of a large number of steps.

  It is useful to express $G_+(x,x_0,t)$ in terms of rescaled variables as in Eq. (\ref{def_rescaled}): 
\begin{eqnarray}
 &&  G_+(x,x_0,t)=\frac{1}{t^{1/\alpha}} Z(y, y_0) , \label{scalingrelation}
\end{eqnarray}
The scaling function $Z(y,y_0)$ depends explicitly on $\alpha$ and we compute it here in perturbation theory for $\epsilon=2-\alpha \ll~1$:
\begin{eqnarray}
Z(y,y_0)= Z^{(0)}(y,y_0) - \epsilon \, Z^{(1)} (y, y_0) + {\cal O}(\epsilon^2) \:. 
\end{eqnarray}

In presence of an absorbing boundary at the origin, the action of the fractional operator can be  written as
\begin{equation}\label{eigensystem2b}
\int_{0}^{\infty} \, d x'  \psi_q(x')    \int_{- \infty}^{\infty} \, \frac{d k}{2 \pi}  (- |k|^\alpha) \, e^{i k (x-x')} =E(q) \psi_q(x),
\end{equation}
and the solution is known only for $\alpha=2$:
\begin{equation}
 \label{discrete_eigen}
\psi_q(x)=\theta(x)\sqrt{\frac{2}{\pi}}\sin(q x) \;, \; E(q)=-q^2 \;,\; q>0 \;,
\end{equation}
where $\theta(x)$ is the Heaviside function: $\theta(x) = 1$ if $x \geq 0$ and $\theta(x) = 0$ if $x <0$.  At zeroth order in $\epsilon$, the scaling function $ Z^{(0)}(y, y_0)$ can be computed from Eq.~(\ref{sol1}) with $\psi_q(x)$ given in Eq.~(\ref{discrete_eigen}). Using the identity $2  \sin(k y) \sin(k y_0)= \cos{[k (y-y_0)]} -\cos{[k(y+y_0)]}$ one obtains: 
\begin{eqnarray}
 Z^{(0)}(y, y_0)&=&\theta(y)\theta(y_0)\int_0^{\infty}\frac{ dk}{\pi}e^{- k^2} \big(\cos[k (y-y_0)] \nonumber \\
 &&-\cos[k(y+y_0)] \Big) \;.
 \end{eqnarray}
Let us note that the same result can be straightforwardly obtained using  the method of images:
\begin{equation}\label{Z0}
 Z^{(0)}(y, y_0)=\theta(y)\theta(y_0)[  R^{(0)}(y-y_0)-R^{(0)}(y+y_0)] \;.
 \end{equation}
At the first order in $\epsilon$ we first compute the matrix element $ \langle  q |{\cal H}_1| q'   \rangle $ which has the form given in Eq.~(\ref{H1}) with the prescription that the integrals over $x_1$ and $x_2$ are performed over the interval $(0,\infty)$ and  $\psi_q(x)$ are the eigenvectors given in Eq.~(\ref{discrete_eigen}). The integrals over $x_1$ and $x_2$ need to be regularized to be well defined and this can be done via the identity:
\begin{eqnarray}
\lim_{\epsilon \to 0} \int_{0}^\infty \frac{dx}{\pi}  e^{i k x- \epsilon x} \sin(q x)  = \text{PV} \frac{q}{\pi (k^2-q^2)} + \nonumber\\ +\frac{i}{2} \left( \delta(q-k)-\delta(q+k)\right) \;,
\end{eqnarray}
where $\text{PV}$ indicates a principal value. After some algebra, left in the Appendix \ref{app_pv}, one gets [see Eq. (\ref{final_expr})]: 
\begin{equation}\label{matrix_element}
 \langle  q |{\cal H}_1| q'   \rangle=  \delta(q-q') q^2 \log|q|+ \frac{q q'}{2 (q+q')}.
\end{equation}
Combining the latter equation with Eq.~(\ref{sol2_bis}) we can write an expression for $P_+^{(1)}(x,x_0,t)$. Analogously to the case of the  propagator in absence of boundaries,  the integrals involved in the expression of  $P_+^{(1)}(x,x_0,t)$ can be naturally recast in term of the variables $z=x/\sqrt{t}$ and $z_0=x_0/\sqrt{t}$ instead of the correct scaling variables $y$ and $y_0$. Following the same lines of the previous discussion we easily write the scaling function $Z^{(1)}(y,y_0)$ as:
\begin{eqnarray}
\label{sol3}
Z^{(1)}(y,y_0)&=& Z_A(y,y_0)+Z_B(y,y_0) \;, \\
Z_A(y,y_0)&=&R^{(1)}(y-y_0)-R^{(1)}(y_0+y) \;, \label{images1} \\
Z_B(y,y_0)&=&   \iint  d k_1  \, d k_2  \frac{k_1 k_2 (e^{-k_2^2}- e^{-k_1^2})}{\pi (k_1+k_2)(k_2^2-k_1^2)}  \nonumber \\
&&\times \sin(k_1 y) \sin(k_2 y_0) \label{zb} \;,
\end{eqnarray}
where the integrals over $k_1,k_2$ run over the interval $(0,\infty)$.  
It is worth to stress that the term $Z_A$ corresponds to the images method prediction while the term $Z_B$ represents the violation of the images prediction at the first order level.

It is easy to realize that the probability density function $R_+(y)$ is simply related to $Z(y,y_0)$ in the following way
\begin{equation}\label{ratio_r+}
R_+^{}(y)=  \lim_{y_0 \to 0} \frac{Z(y,y_0)}{\int_{0}^\infty   \, dy'\,   Z(y',y_0)} \;.
\end{equation}
For $\alpha=2$ one has from Eq. (\ref{Z0}) in the limit $y_0 \to 0$:
\begin{equation}\label{Z0_smallw}
Z^{(0)}(y,y_0) = y_0 \tilde Z^{(0)}(y) + {\cal O}(y_0^2)\;, \; \tilde Z^{(0)}(y) = \frac{y}{2 \sqrt{\pi}} e^{-\frac{y^2}{4}}  \;,
\end{equation}
which yields 
\begin{equation}\label{Rp0}
R_+^{(0)}(y)=  \frac{y}{2} \, e^{-\frac{y^2}{4}} \;.
\end{equation}

The integrals in Eqs. (\ref{images1}) and (\ref{zb}) which give  the term $R_+^{(1)}(y)$ have to be discussed carefully and the details are left in the Appendix \ref{app_propag}. The net result of this analysis is that $R_+(y)$ can be written as
\begin{eqnarray}
&&R_+(y) = R_+^{(0)}(y)  \left(1 + \epsilon W_+(y) + {\cal O}(\epsilon^2)\right) \;,
\end{eqnarray}
where $W_+(y)$ can be expressed in terms of elementary and special functions, see Eq. (\ref{W_elementary}). From this expression, one obtains the asymptotic behaviors of $R_+(y)$. In the limit $y \to 0$, one finds
\begin{equation}
R_+(y) \sim \frac{y}{2}  \ -\frac{\epsilon \, y}{4}(\log y + \kappa)   + {\cal O}(y^2 \log{y}, \epsilon^2) \;,    \label{rplus_smallx}
\end{equation}
where $\kappa=2-\log{2} - \frac{3}{2}\gamma_E$. In particular, the small $y$ behavior in Eq.~(\ref{rplus_smallx}) is consistent with $R_+(y) \sim \, y^{\frac{\alpha}{2}} $ in agreement with previous findings \cite{ZK95, ZRK07}. For $y \to \infty$ one finds
\begin{equation}
R_+(y) \sim \epsilon \left(\frac{2}{y^3} + \frac{3 \sqrt{\pi}}{y^4} + \frac{32}{y^5} \right) + {\cal O}(y^{-6}, \epsilon^2)  \;,\label{rplus_largex}
\end{equation}
where the leading term, vanishing as $1/y^3$, is expected from previous analysis~\cite{ZK95}. We notice that the right tail of
the $R_+(y)$ has the same behavior as the right tail of $R(y)$. Quite interestingly, our perturbative result shows that
\begin{eqnarray}\label{ratio_perturb}
\frac{c_+}{c} = \lim_{y \to \infty} \frac{R_+(y)}{R(y)} = 2 + {\cal O}(\epsilon) \;,
\end{eqnarray}
where $c$ and $c_+$ are defined in Eq. (\ref{R_intro}) and Eq. (\ref{result_large}) respectively. Another signature of the boundary, revealed by this perturbative calculation, appears in the sub-leading correction which vanishes as $1/y^4$ for $R_+(y)$ [see Eq. (\ref{rplus_largex})] instead of $1/y^5$, as for $R(y)$ [see  Eq. (\ref{free_asymp})].

\subsection{Discussion and conjectures}

It is interesting to compare our perturbative results, valid in principle for $2 - \alpha \ll 1$, with the exact results which we can obtain for 
the special case $\alpha = 1$. In this case, corresponding to Cauchy random variables, {\it i.e.} $\varphi(\eta) = \pi^{-1}(1+\eta^2)^{-1}$, one can use the results which were obtained by Darling \cite{DD56} and Nevzorov \cite{VN82} in the context of the extreme statistics of such L\'evy statistics, to obtain an exact result for $R_+(z)$ in terms of a single integral:  
\begin{eqnarray}
&&R_+(z) = - \sqrt{z} \int_0^1 g\left(\frac{z}{v} \right) v^{-3/2} (1-v)^{-1/2} \, dv  \; \label{g_Cauchy} \\
&&g(z)  = \frac{d}{dz} \left[\frac{1}{\pi} \frac{1}{(1+z^2)^{3/4}} \exp{\left(-\frac{1}{\pi} \int_0^z \frac{\log{u}}{1+u^2} du \right)} \right] \;. \nonumber
\end{eqnarray} 
Its asymptotic behaviors are given by
\begin{eqnarray}
&&R_+(z) \sim \frac{1}{2} \sqrt{z}  \;, z \to 0\;\\
&&R_+(z) \sim \frac{2}{\pi} \frac{1}{z^2} + \frac{16}{6 \pi^{2}} \frac{\log{z}}{z^3} + {\cal O}(z^{-3})    \;, \; z \to \infty\;. \label{asympt_cauchy}
\end{eqnarray} 
On the other hand, one has in this case~(\ref{free_levy_intro}) $R(y) \sim 1/(\pi y^2)$ when $y \to \infty$ such that one obtains also in this case $\alpha = 1$
\begin{eqnarray}\label{ratio_exact}
\frac{c_+}{c} = \lim_{y \to \infty} \frac{R_+(y)}{R(y)} = 2 \;.
\end{eqnarray}
Based on the perturbative result obtained above (\ref{ratio_perturb}) and on this exact result (\ref{ratio_exact}) we conjecture that this relation $c_+ = 2 c$ actually holds for all value of $\alpha$, as stated in the introduction (\ref{relation_c}). This conjecture is corroborated below by our numerical simulations. 

Besides, we interpret the exponent of the sub-leading correction in $R_+(y)$, which decays as $1/y^{4}$ in Eq. (\ref{rplus_largex}), as $4 = 2 + \alpha$, with $\alpha = 2 + {\cal O}(\epsilon)$ while the sub-leading corrections in $R(y)$ decay as $1/y^5$ (\ref{free_levy_intro}) with, instead $5 = 1 + 2 \alpha$, for $\alpha = 2 + {\cal O}(\epsilon)$. This leads us to conjecture that the subleading corrections behave actually differently for $\alpha > 1$ or $\alpha < 1$, as announced in the introduction (\ref{result_large}) 
\begin{eqnarray}\label{conjecture}
R_+(y) = \frac{c_+}{y^{1+\alpha}} \; + \; 
\begin{cases}
&\frac{d_+}{y^{2+\alpha}} + o (y^{-2-\alpha})\;, \; 2 > \alpha > 1 \;, \\
& \\
&\frac{d_+}{y^{1+2\alpha}} + o(y^{-1-2 \alpha}) \;,\; 1 > \alpha > 0 \;.
\end{cases}
\end{eqnarray}  
Hence $\alpha = 1$ appears as a critical value regarding these sub-leading corrections, for which it is not surprising to observe logarithmic corrections (\ref{asympt_cauchy}). This also implies that the coefficient $d_+$ above (\ref{conjecture}) is diverging when $\alpha \to 1$ from above. Below, we will test this behavior by means of numerical simulations.

\section{Numerical simulations}

\begin{figure}
   \includegraphics[width=8.75cm]{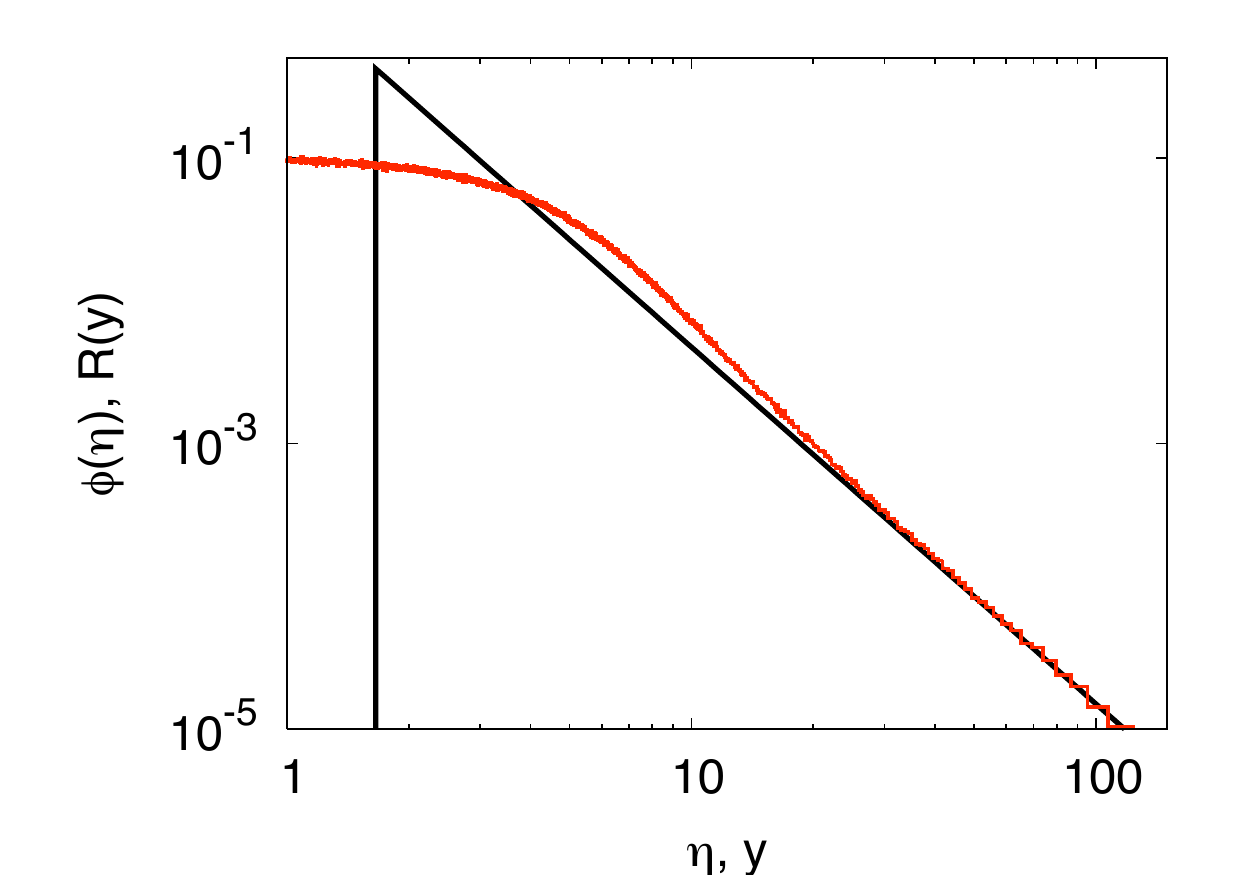} 
\caption{Case $\alpha=1.5$. Behavior of the right tail 
for $\phi(\eta)$ (Black)  and the the rescaled final position, $y$, of free random walks of $n=1000$ step (Red). Histograms are performed using $10^7$ samples. } 
\label{f:1}
\end{figure} 
 
\begin{figure}
\includegraphics[width=8.75cm]{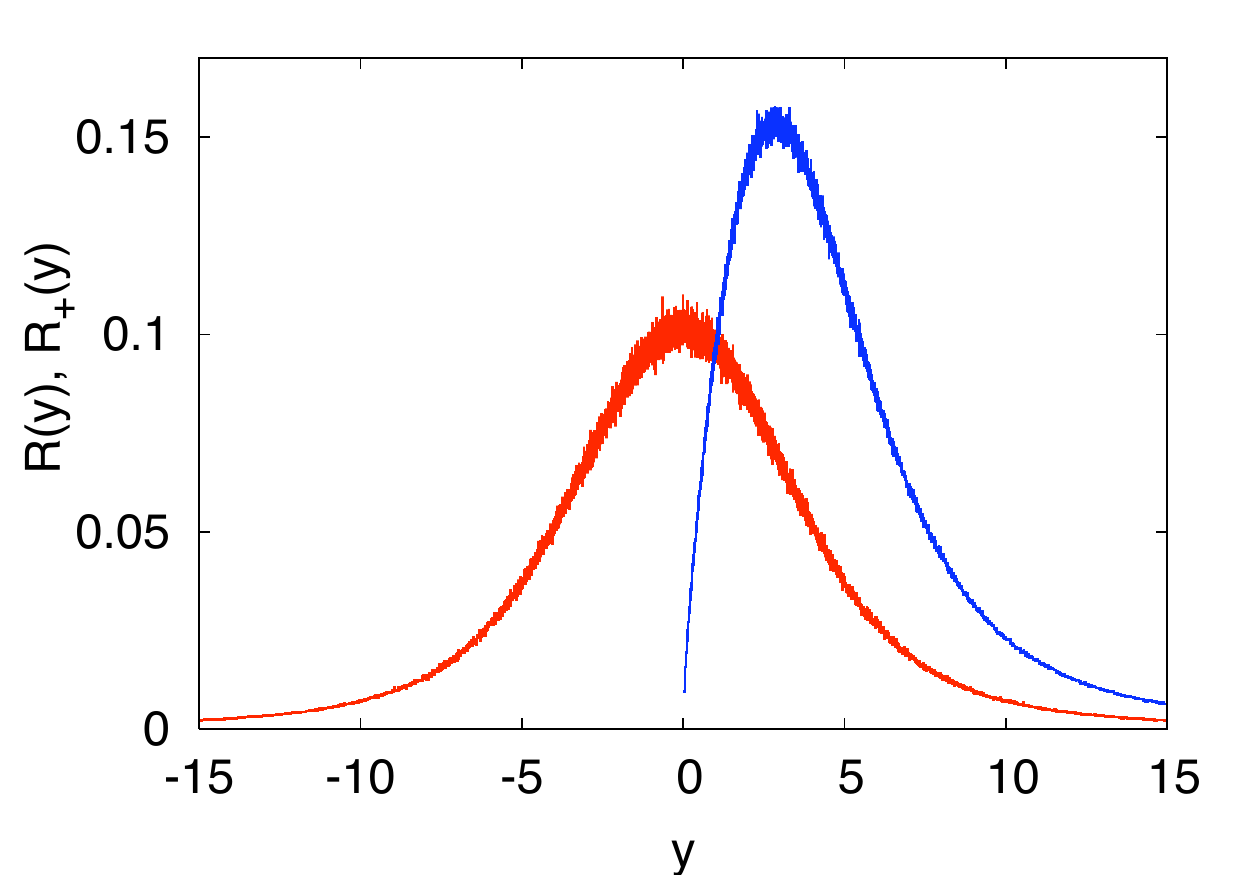}   
\caption{ Case $\alpha=1.5$.  Free random walks of $n=1000$ steps (Red) and random walks constrained to be positive (Blue). Histogram are performed using $10^8$ samples. } 
\label{f:2}
\end{figure}

\begin{figure}
\includegraphics[width=8.75cm]{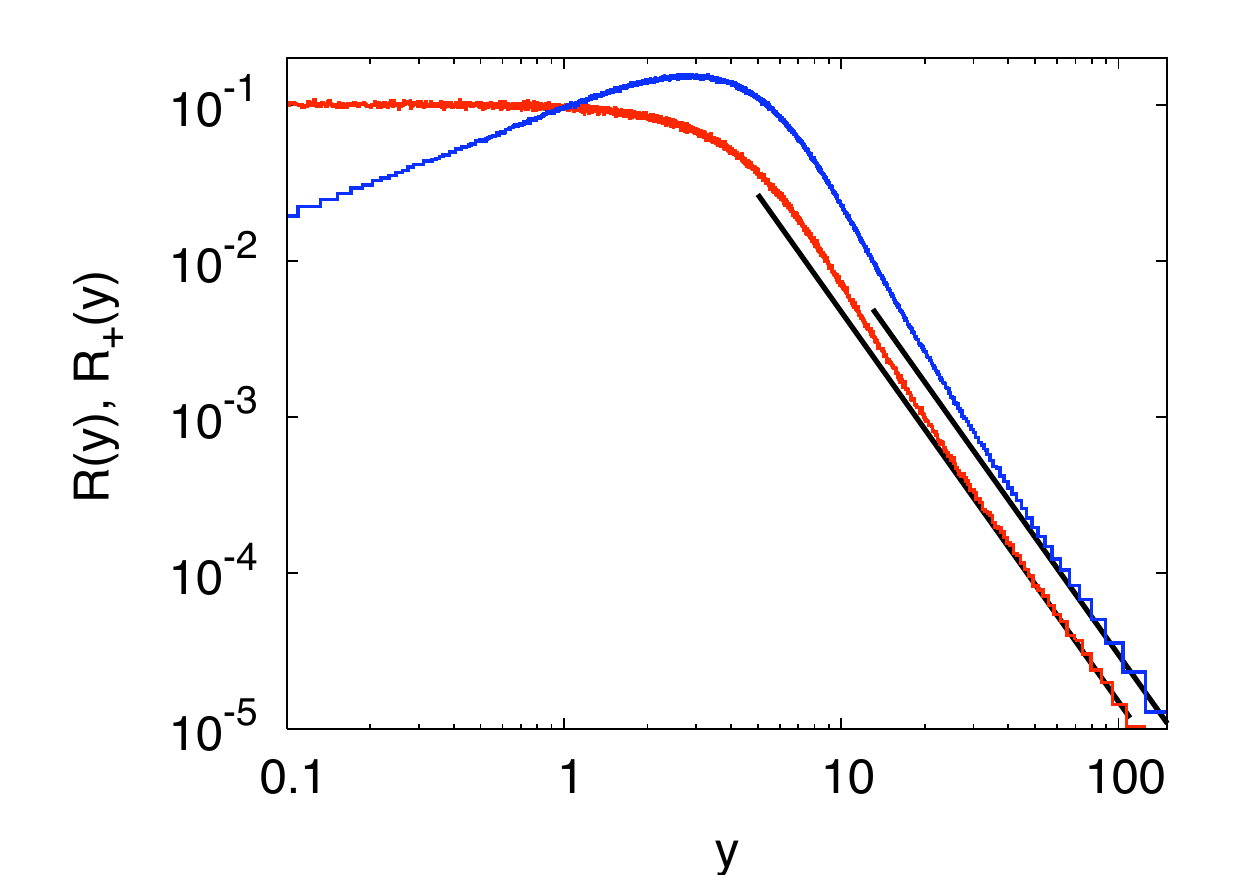}   
\caption{Case $\alpha=1.5$.  Free random walks of $n=1000$ steps (Red) and random walks constrained to be positive (Blue). Study of the tails. Histogram are performed using $10^8$ samples.  The expected tails: $1.5/y^{5/2}$ and $3/y^{5/2}$ are also drawn (Solid Line). } 
\label{f:3}
\end{figure}

We consider the case where the increments distribution $\phi(\eta)$ is the symmetric Pareto distribution:
\begin{eqnarray}\label{def_pareto}
\phi(\eta)=  \begin{cases}    &\frac{\alpha}{|\eta|^{\alpha+1}}\;, {\rm for} \; |\eta| > 2 ^{\frac{1}{\alpha}}\\   & 0 \;, {\rm otherwise}.   \end{cases}
\end{eqnarray}
This distribution can be sampled efficiently using random number drawn from a uniform distribution:
\begin{eqnarray}\label{pareto}
\eta=  \begin{cases}    &   \left[ \rm{ran}(0,\frac{1}{2})\right]^{-\frac{1}{\alpha}}     \;, {\rm with}\, {\rm probability } \; \frac{1}{2}\\  &  - \left[ \rm{ran}(0,\frac{1}{2})\right]^{-\frac{1}{\alpha}} \;, {\rm with} \, {\rm probability}   \; \frac{1}{2};   \end{cases}
\end{eqnarray}
where $\rm{ran}(0,\frac{1}{2})$ is a random number in the interval $(0,\frac{1}{2})$.  We construct a large number of random walks, for each random walk we record the final position $x(n)$ after $n$ steps and compute the correspondent  rescaled variable $y=x_n/n^{1/\alpha}$. We first present our data for $\alpha=1.5$, for which $\phi(\eta) = 1.5/ |\eta|^{5/2}$ for $|\eta|>2^{2/3}$.   For large $n$ the distribution of $y$ should converge to the a stable distribution centered around $0$ and with an asymptotic tail
\begin{eqnarray}
R(y) = \frac{3}{ 2 \, y^{5/2}} +\frac{24}{y^4} + o(y^{-4}) \;.
\end{eqnarray}
This prediction is confirmed by our direct simulation, in Fig. \ref{f:1} we show that the tail of $\phi(\eta)$ and $R(y)$ coincide when $\eta, y \to \infty$. The symmetric distribution $R(y)$ is also plotted in Fig. \ref{f:2} where we also show  $R_+(y)$, the histogram of the rescaled final position of the random walks constrained to be positive.   $R_+(y)$ is clearly defined only for positive $y$, vanishes at $y=0$ and, when $y \to \infty$, decays as $c_+/y^{5/2}$. One of our main prediction is that, for large $y$, $R_+(y)/R(y)=2$ which means   $c_+=3$ for our model. This is verified in Fig.\ref{f:3}.

\begin{figure}[t]
\includegraphics[width=8.75cm]{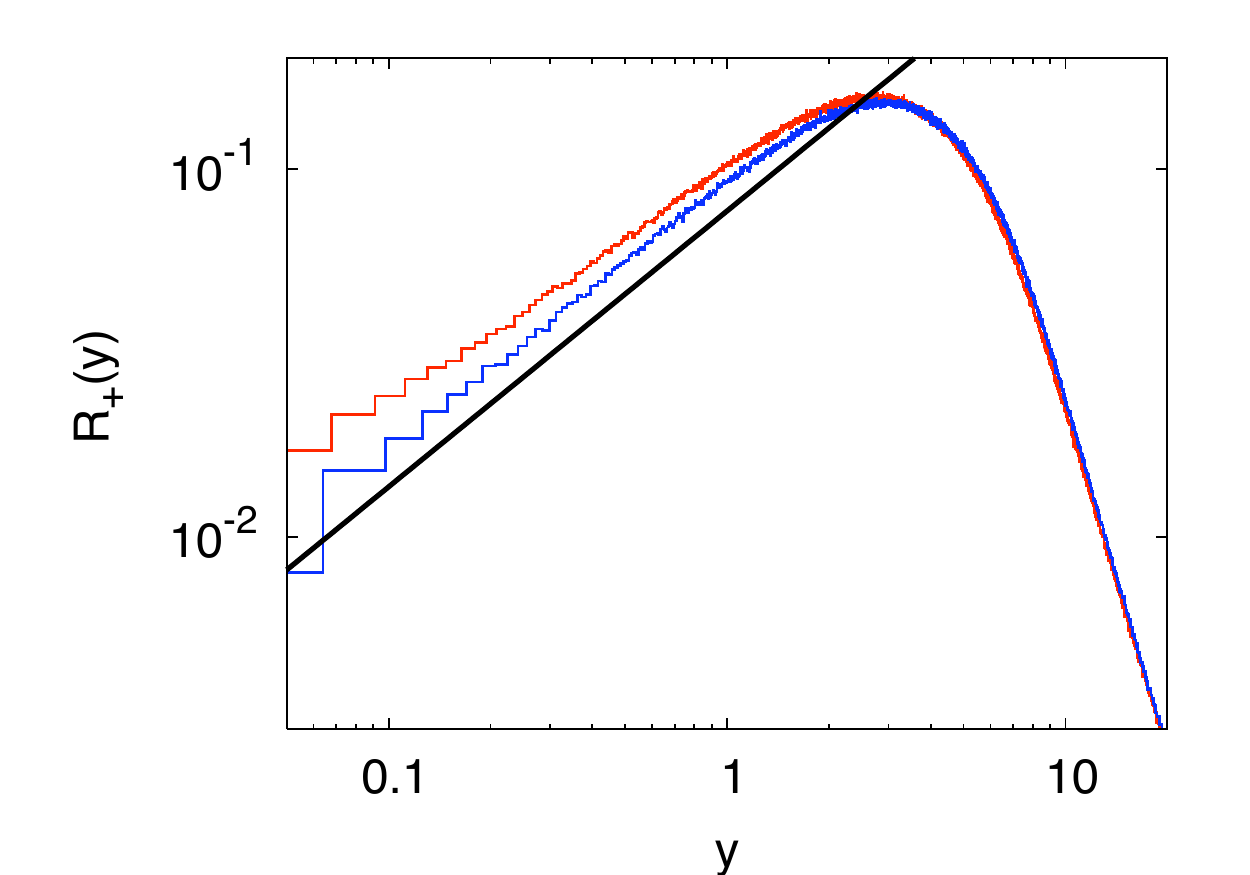}   
\caption{ Case $\alpha=1.5$. Finite size effects for random walks constrained to be positive.  Long random walks ($n=5000$ -Blue)  vs short ranomd walks ($n=250$ - Red). Histograms are perfomed over $10^7$ samples. The slope  $y^{\frac{\alpha}{2}}$ is plotted as a guide of the eye (Black line).  } 
\label{f:4}
\end{figure}

\subsection{Finite size effect and different values of $\alpha$}

\begin{figure}[t]
\includegraphics[width=8.75cm]{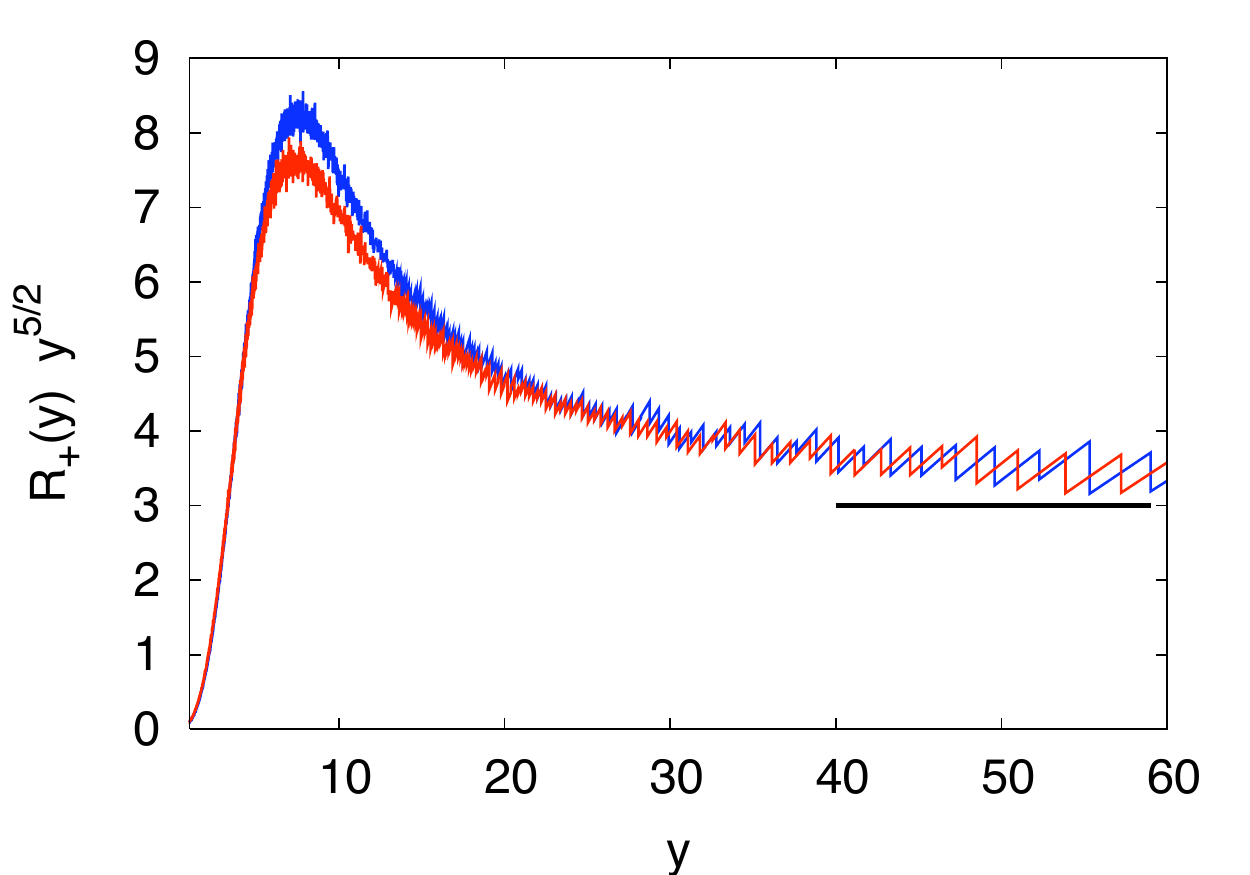}   
\caption{ Case $\alpha=1.5$. Finite size effects for random walks constrained to be positive.  Long random walks ($n=5000$ -Blue)  vs short ranomd walks ($n=250$ - Red). The constant $c_+=3$ is recovered. } 
\label{f:5}
\end{figure}

Stable distributions and universal behavior are expected  in the limit of a large number of steps ({\it i.e.} $n \to \infty$). In our numerical simulation the asymptotic behavior of $R(y)$ and $R_+(y)$ is studied for $n=1000$. Is this number enough to capture universality? In Fig.~\ref{f:4} and Fig.~\ref{f:5} we study how the finite number of steps affect the function $R_+(y)$.  Finite size effects are visible close to the boundary $y=0$ where, only for very large size, the distribution vanishes with the predicted exponent  $\alpha/2=0.75$. For $y\simeq 10$  the convergence with the size $n$ becomes faster and the constant
\begin{equation}
c_+=\lim_{y \to \infty} R_+(y) \, y^{\alpha+1}
\end{equation}
 can be correctly estimated even with a moderate number of steps (see Fig.\ref{f:5}).

\begin{figure}[b]
\includegraphics[width=8.75cm]{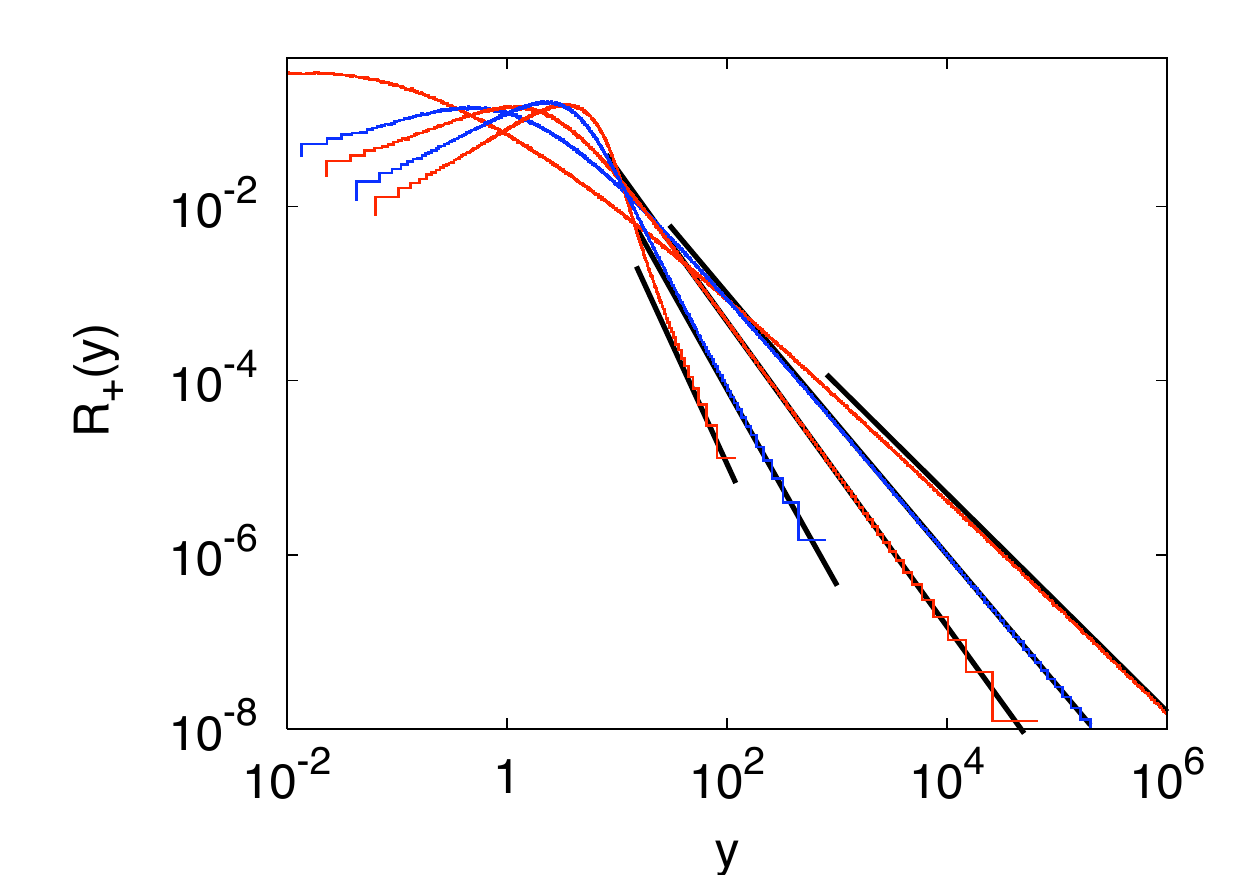}   
\caption{  Random walks constrained to be positive for $n=1000$ steps and  $\alpha=1.75, 1.25, 0.75,0.5, 0.25$ (from left to right). Histograms are perfomed over $10^7$ samples. The asymptotic behavior $c_+/y^{\alpha+1}$ is also plotted (Black line).  } 
\label{f:6}
\end{figure}

Finally we have checked that our result for $c_+$ apply to all range of $0<\alpha<2$ for symmetric L\'evy flights. The asymptotic tail  is more and more pronounced as $\alpha \ll 2$. This means that the insights given by our perturbative calculation are actually valid for all L\'evy flights (see Fig.\ref{f:6}).

\subsection{Subleading corrections}
\begin{figure}[ht]
\includegraphics[width=8.75cm]{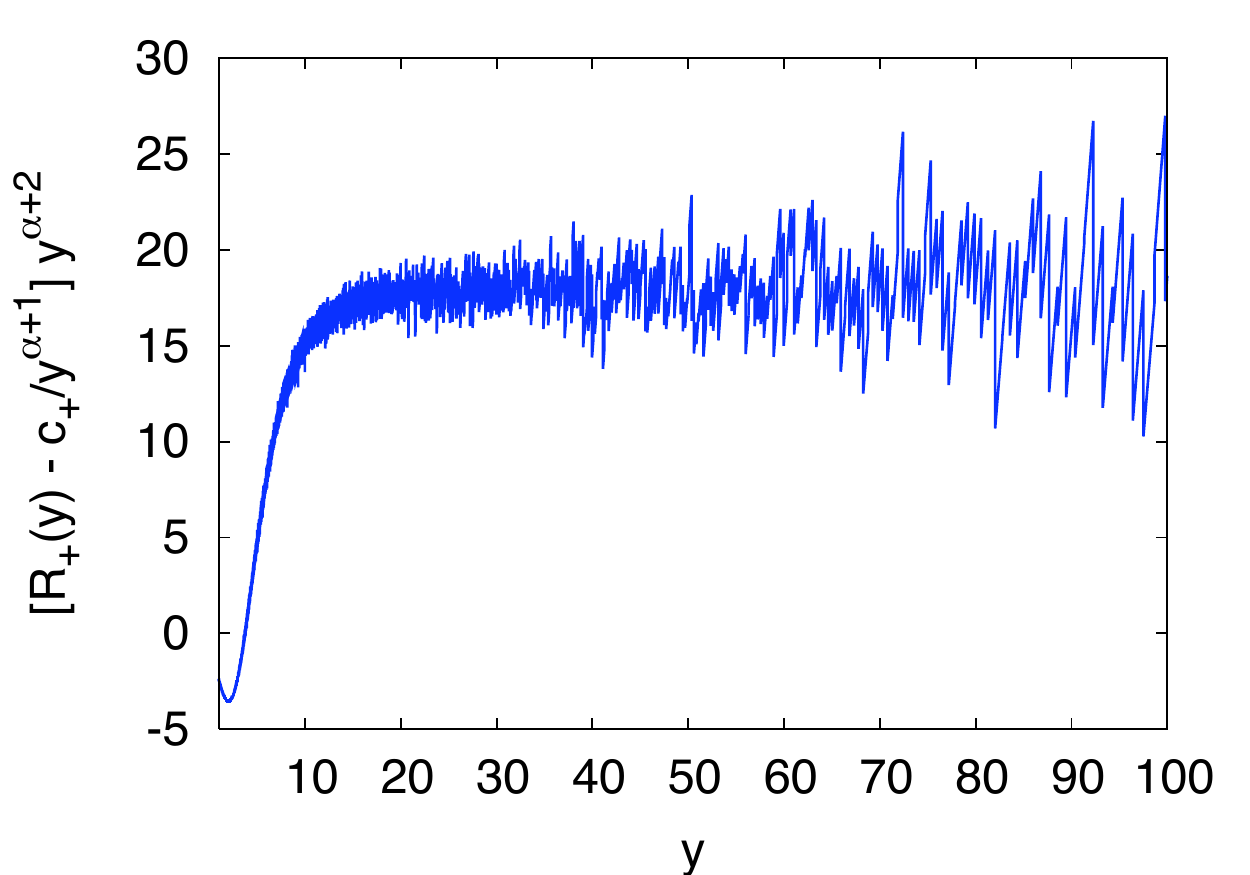}   
\caption{  Sub-leading correction for $\alpha=1.25>1$. Random walks constrained to be positive for $n=1000$ steps. Histograms are performed over $10^8$ samples. The  plateau reached at large $y$ gives a numerical estimation of $d_+$. } 
\label{sub}
\end{figure}

\begin{figure}[h]
\includegraphics[width=8.75cm]{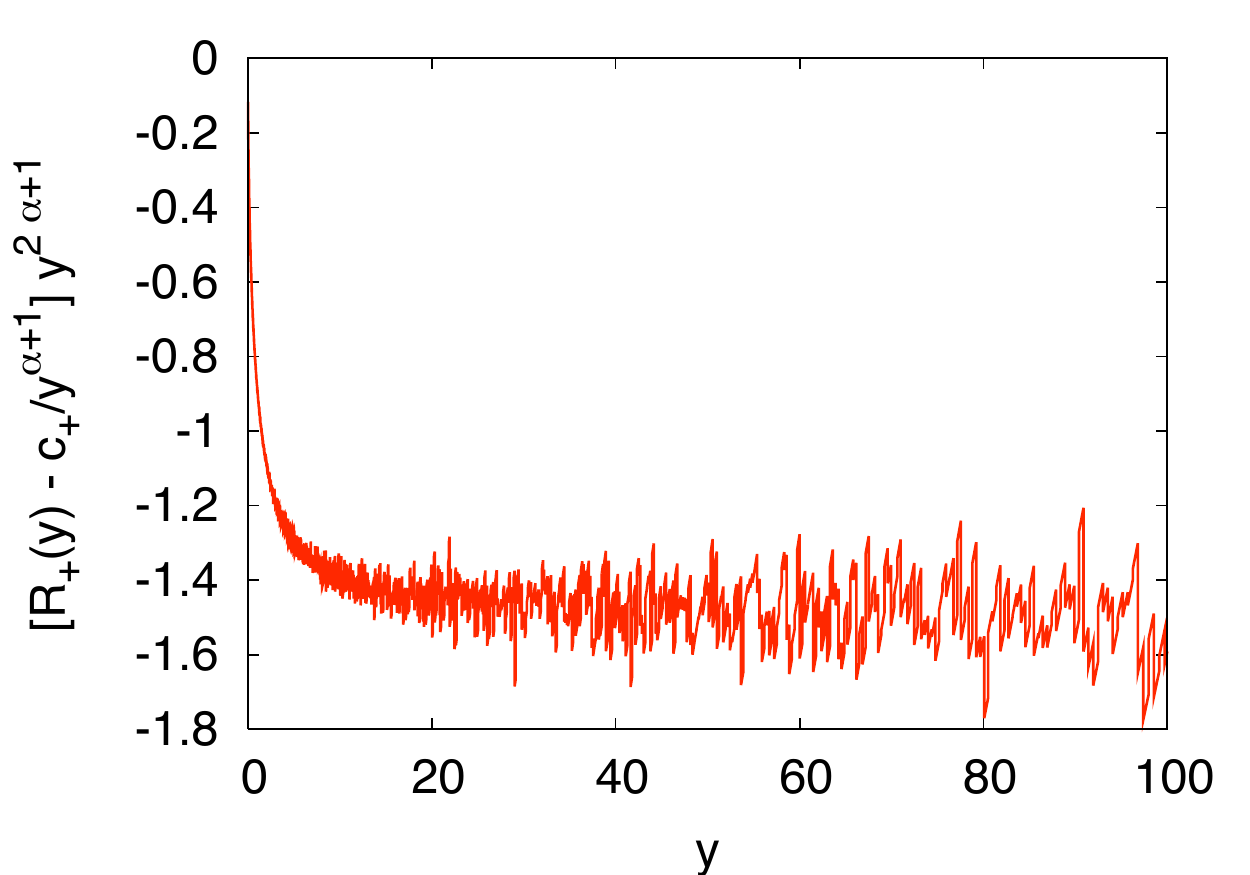}   
\caption{  Sub-leading correction for $\alpha=0.5 <1$. Random walks constrained to be positive for $n=1000$ steps. Histograms are performed over $10^7$ samples. The  plateau reached at large $y$ gives a numerical estimation of $d_+$. } 
\label{sub2}
\end{figure}

We also check the behavior of the sub-leading correction~(\ref{conjecture}). For $\alpha >1$ this correction is expected to behave like $d_+ y^{-\alpha-1}$, while for $\alpha<1$ we expect that it decays as  $\sim d_+ y^{-2 \alpha-1}$. This prediction is confirmed in Fig.~\ref{sub} for $\alpha>1$ and Fig.~\ref{sub2} for $\alpha<1$.

\section{Conclusion}

To conclude, we have presented a perturbative approach to the study of a L\'evy flight, of index $0 < \alpha < 2$ on a half-line, where the perturbative parameter is $\epsilon = 2 - \alpha$. This approach, following the work of Zoia, Rosso and Kardar \cite{ZRK07},  amounts to construct a perturbative solution of the fractional Fokker-Planck equation (\ref{fp_intro}) with appropriate (non-local) boundary conditions. Here, at variance with Ref. \cite{ZRK07} the perturbation theory is carried out directly for a process which is continuous both in space and time. 

We have then used this perturbative method to compute, to order ${\cal O}(\epsilon)$, the  probability density function, $R_+$, of the position of such a walker with an absorbing wall at the origin. A different perturbation scheme (based on path integral) was used recently \cite{WMR11} to compute the same quantity for the fractional Brownian motion, a non-Markovian process displaying anomalous diffusion. Our main result here  is to give a precise conjecture, valid for any value of $\alpha$, on the relation between  the tail of this distribution and  the tail of the steps  of the random walk. Numerical simulation confirm our conjecture.

This perturbative scheme opens the way to an analytical study and can be used for any confined domain for which the Brownian solution, corresponding to $\alpha = 2$, is known. More realistic confined geometry, like the one relevant to the scattering of solar photons, can be studied along these lines. Here we have proposed the simplest of these geometries. A first extension of the present study concerns the study of the extreme statistics of a L\'evy bridge, which is a L\'evy random walk on the time interval $[0,t]$ constrained to start and end at the origin. Such a constrained L\'evy random walk, for which little is known, has recently received some attention in statistical physics~\cite{schehr_levyarea} -- in relation with some real-space condensation phenomena~--, as well as in finance \cite{levybridge_finance}. Another interesting application of such a perturbative calculation could be the study of non-intersecting L\'evy walkers, the so-called "vicious"  L\'evy walkers, which were recently introduced in Ref. \cite{vicious_levy}.    

\begin{acknowledgments}
This work was supported by the France-Argentina MINCYT-ECOS A08E03. R. G.-G. acknowledges the hospitality at LPT and LPTMS in Orsay, A. R. and G. S acknowledge the hospitality at the Centro Atomico in Bariloche where part of this work was done. A. R. acknowledges support by ANR grant 09-BLAN-0097-02. G. S. acknowledges support by ANR grant 2011-BS04-013-01 WALKMAT.  
\end{acknowledgments}

\appendix

\section{Perturbative scheme to order ${\cal O}(\epsilon)$} \label{a_1}

In this appendix we give a short derivation of the expression for the first correction $P^{(1)}(x,x_0, t)$ (\ref{sol2}) given in Eq.~(\ref{sol2_bis}). 
The equation for $P^{(1)}(x,x_0,t)$ reads, see Eqs. (\ref{def_H}, \ref{sol2}):
\begin{equation}\label{a11}
\partial_t P^{(1)}(x,x_0,t)= {\cal H}_0 P^{(1)}(x,x_0,t) + {\cal H}_1 P^{(0)} (x,x_0,t) \;, 
\end{equation}
 with the initial conditions $P^{(1)}(x, x_0, t=0) = 0$ and   $P^{(0)}(x,x_0, t)$ is given in (\ref{sol1}).
 Eq. (\ref{a11}) is  inhomogeneous and we can obtain its solution using 
 the method of variation of constants. We look for a solution under the form
\begin{eqnarray}\label{sola1}
  P^{(1)}(x,x_0,t) = \int c^{(1)}_q(t) \psi_q(x) e^{E(q) t} dq \;,
\end{eqnarray}
where $\psi_q$ and $E(q)$ are the eigenvectors and eigenvalues of $ {\cal H}_0$, while the coefficient $c^{(1)}_q(t)$ depends explicitly on time $t$. Plugging Eqs. (\ref{sola1}) and  (\ref{sol1}) into (\ref{a11}) one finds 
\begin{eqnarray}\label{eq_c1_inter}
  &&\int  \,dq' \,  e^{E(q') t} \psi_{q'} (x) \partial_t c^{(1)}_{q'}(t) \\
  &&=   \int dq'  e^{E(q')t}   \psi_{q'}^*(x_0) { \cal H}_1   \psi_{q'}(x)   \;.
\end{eqnarray}
We  multiply the left and right hand sides of Eq. (\ref{eq_c1_inter}) by $\psi_{q}^*(x)$ and we then integrate both sides over $x$. Using the normalization relation $\int \, d x\, \psi_q^*(x) \psi_{q'}(x)=\delta(q-q')$ we obtain
\begin{eqnarray}
e^{E(q) t}\partial_t c^{(1)}_q(t) = \int dq' \langle q | { \cal H}_1 | q' \rangle \psi_{q'}^*(x_0) e^{E(q') t} \;,
\end{eqnarray}
where we have used the bra-ket notation, borrowed from Quantum Mechanics, and 
such that $\langle x | q \rangle = \psi_q(x)$. This can be straightforwardly integrated, using the boundary condition at time $t = 0$ 
\begin{equation} \label{a1last}
c^{(1)}_q(t) = \int dq' \langle q | {\cal H}_1 | q' \rangle \psi_{q'}^*(x_0) \frac{e^{E(q') t}-e^{E(q) t}}{E(q) - E(q')} \;.
\end{equation}
Combining (\ref{a1last}) and (\ref{sola1}) one obtains Eq.~(\ref{sol2}) given in the text.

\section{Evaluation of an integral}\label{app_pv}

Here we compute the principal value of the integral entering the matrix element in Eq. (\ref{matrix_element}):
\begin{eqnarray}\label{structure}
I(q,q') &=& \text{PV} \int_{-\infty}^{\infty} d k \frac{ q q' k^2 \ln|k|}{\pi^2 (k^2-q^2)(k^2-q'^2)} \nonumber \\
&=& \frac{2qq'}{\pi^2}\text{PV}\int_0^\infty dk \frac{k^2 \ln k}{(k^2-q^2)(k^2-q'^2)} \;.
\end{eqnarray}
Without any loss of generality, since $I(q,q') = I(q',q)$, we assume $q'\geq q$. Performing the change of variable $k = s\,q$, one obtains straightforwardly
\begin{eqnarray}
I(q,q') = \frac{2q'}{\pi^2} f_1\left(\frac{q'}{q}\right) \ln{q} + \frac{2q'}{\pi^2} f_2\left(\frac{q'}{q}\right) \;, 
\end{eqnarray}
where
\begin{eqnarray}\label{def_f1_f2}
&&f_1(x) = \text{PV} \int_0^\infty ds \frac{s^2}{(s^2-1)(s^2-x^2)} \;, \\
&&f_2(x) = \text{PV} \int_0^\infty ds \frac{s^2 \ln s}{(s^2-1)(s^2-x^2)} \;.
\end{eqnarray}
We notice that $f_2(x)$ can be written as
\begin{equation}\label{trick}
f_2(x) = f_1(x) \ln x + \text{PV}\int_0^\infty ds \, s^2 \frac{\ln(s/x)}{(s^2 - 1)(s^2-x^2)} \;.
\end{equation}
The purpose of this trick (\ref{trick}) is that the integrand has now only a simple pole in $s=1$ so that, taking the principal value, yields a perfectly smooth
function of $x$. This integral can then be evaluated to yield:
\begin{eqnarray}
f_2(x) = f_1(x) \ln x + \frac{\pi^2}{4(1+x)} \;, 
\end{eqnarray}
so that one has
\begin{eqnarray}
I(q,q') = \frac{2q'}{\pi^2} f_1\left(\frac{q'}{q}\right) \ln{q'} + \frac{q\,q'}{2(q+q')} \;.
\end{eqnarray}
Looking at the expression of $f_1(x)$ above (\ref{def_f1_f2}), one sees that for $x=1$, the integrand has a double pole in $s=1$ and one thus expects $f_1(x)$ to be highly singular in $x=1$. To characterize it, we compute its Fourier transform
\begin{eqnarray}
\hat f_1(p) = 2 \int_0^\infty dx f_1(x) \cos{(px)} \;,
\end{eqnarray} 
which yields after straightforward manipulations of Eq. (\ref{def_f1_f2}) (which can be done e.g. with Mathematica) 
\begin{eqnarray}
\hat f_1(p) = \frac{\pi^2}{2} \cos{p} \;,
\end{eqnarray}
which yields 
\begin{eqnarray}
f_1\left(\frac{q'}{q}\right) = \frac{\pi^2 q}{4} \delta(q-q') \;,
\end{eqnarray}
and finally
\begin{eqnarray}\label{final_expr}
I(q,q') = \delta(q-q') \frac{1}{2} q^2 \ln q + \frac{q\,q'}{2(q+q')} \;,
\end{eqnarray}
which is used in the Eq.~(\ref{matrix_element}) in the text.

\section{Details about the perturbative calculation of the propagator with an absorbing wall}\label{app_propag}

This appendix is devoted to the analysis of $Z^{(1)}(y,y_0)$ given by the sum of the two terms in Eqs. (\ref{sol3}, \ref{images1}, \ref{zb})
in the limit $y_0 \to 0$. The first term $Z_A(y,y_0)$ is easy to analyse and yields
\begin{eqnarray}\label{za_small_w}
Z_A(y,y_0) = -2 y_0 \partial_y R^{(1)}(y) + {\cal O}(y_0^2) \;,
\end{eqnarray}
where $R^{(1)}(y)$ admits the integral representation given in Eq. (\ref{firstintb}). 

The analysis of the small $y_0$ behavior of $Z_B(y,y_0)$ given in Eq. (\ref{zb}) is more subtle. To deal with this double integral over $k_1$ and $k_2$, we first make the change of variable $k_1 = u'$ and $k_2 = u u'$ and observe that the integral over $u'$ can then be performed to yield
%
\begin{eqnarray}
&&Z_B(y,y_0) = \frac{1}{4\sqrt{\pi}} \int_0^\infty du \frac{u}{(u-1)(u+1)^2} \Bigg[e^{-\frac{(y+u y_0)^2}{4}} \nonumber \\
&&- e^{-\frac{(y-u y_0)^2}{4}}  + \frac{1}{u} \left(e^{-\frac{(y-u y_0)^2}{4u^2}} - e^{-\frac{(y+u y_0)^2}{4u^2}} \right)  \Bigg ] \;.
\end{eqnarray}
Using now the identity
\begin{eqnarray}
\frac{u}{(u-1)(u+1)^2} = \frac{1}{(u+1)^2} + \frac{1}{(u-1)(u+1)^2} \;,
\end{eqnarray}
we split $Z_B(y,y_0)$ into two parts:
\begin{eqnarray}\label{start_zb_app}
Z_B(y,y_0) = \frac{1}{4 \sqrt{\pi}} ({\cal I}_1(y,y_0) + {\cal I}_2(y,y_0)) \;,
\end{eqnarray}
where
\begin{eqnarray}
{\cal I}_1(y,y_0) &=& \int_0^\infty \frac{du}{(u+1)^2} \Bigg[e^{-\frac{(y+uy_0)^2}{4}} - e^{-\frac{(y-u y_0)^2}{4}} \nonumber \\
&+& \frac{1}{u} \left(e^{-\frac{(y-u y_0)^2}{4u^2}} - e^{-\frac{(y+u y_0)^2}{4u^2}} \right)  \Bigg] \;, \label{def_I1}\\
{\cal I}_2(z,w) &=& \int_0^\infty \frac{du}{(u-1)(u+1)^2} \Bigg[e^{-\frac{(y+u y_0)^2}{4}} - e^{-\frac{(y-u y_0)^2}{4}} \nonumber \\
&+& \frac{1}{u} \left(e^{-\frac{(y-u y_0)^2}{4u^2}} - e^{-\frac{(u+ u y_0)^2}{4u^2}} \right)  \Bigg] \;. \label{def_I2}
\end{eqnarray}
In ${\cal I}_2(y,y_0)$ (\ref{def_I2}) the small $y_0$ limit can be taken easily. It yields 
\begin{eqnarray}\label{I2small}
{\cal I}_2(y,y_0) &=& y\,y_0 \int_0^\infty \frac{du}{(u-1)(u+1)^2} \left(\frac{1}{u^2}e^{-\frac{y^2}{4u^2}} - u e^{-\frac{y^2}{4}} \right) \nonumber \\
&+& {\cal O}(y_0^2) \;.
\end{eqnarray}
We now decompose ${\cal I}_1(y,y_0)$ (\ref{def_I2}) into two parts
\begin{eqnarray}
&&{\cal I}_1(y,y_0) = {\cal I}_{11}(y,y_0) + {\cal I}_{12}(y,y_0) \;, \\
&&{\cal I}_{11}(y,y_0) = \int_0^\infty \frac{du}{(u+1)^2} \left(e^{-\frac{(y+u y_0)^2}{4}} - e^{-\frac{(y-u y_0)^2}{4}}\right) \;,  \nonumber \\
&&{\cal I}_{12}(y,y_0) = \int_0^\infty \frac{du}{u(u+1)^2} \left(e^{-\frac{(y-u y_0)^2}{4u^2}} - e^{-\frac{(y+u y_0)^2}{4u^2}}\right) \nonumber \;.
\end{eqnarray}
In ${\cal I}_{12}(y,y_0)$, it is straightforward to obtain the small $y_0$ behavior as
\begin{equation}
{\cal I}_{12}(y,y_0) = y\, y_0 \int_0^\infty \frac{du}{u^2(u+1)^2} e^{-\frac{y^2}{4u^2}} + {\cal O}(y_0^2) \label{I12small} \;.
\end{equation}
The integral in ${\cal I}_{11}(y,y_0)$ contains a logarithmic singularity when $y_0 \to 0$, which is a bit tricky to extract. To do so, we first perform a change of variable $s = y_0 \, u$ and then add and substract the term $y_0\int_0^\infty ds (e^{-y^2/4} - e^{-y^2/4- y s})/(s+y_0)^2$. 
Now using
\begin{eqnarray}\label{log_sing}
\int_0^\infty ds \frac{1}{(s+y_0)^2}\left(e^{-y s} - 1 \right) &=& y (\log{y_0} + \log{y} + \gamma_E) \nonumber \\
&+& {\cal O}(y_0 \log y_0) \;,
\end{eqnarray}
one obtains, for $y_0 \to 0$:
\begin{eqnarray}\label{final_I11}
{\cal I}_{11}(y,y_0) &=& (y_0\log{y_0}) \, y e^{-\frac{y^2}{4}} + y_0 y e^{-y^2/4}\left(\gamma_E + \log{y}\right) \nonumber \\
&-& y_0 e^{-y^2/4} {\cal Q}(y) +{\cal O}(y_0^2 \log y_0) \;, 
\end{eqnarray}
where
\begin{equation}\label{expr_qz}
{\cal Q}(y) = \int_0^\infty \frac{ds}{s^2} \left(e^{-y s} - 1 + 2 \, e^{-\frac{s^2}{4}} \sinh{\left(\frac{y\,s}{2} \right)} \right) \;, 
\end{equation}
Finally combining Eqs. (\ref{I2small}), (\ref{I12small}) and (\ref{final_I11}) one obtains the small $y_0$ behavior of $Z_B(y,y_0)$ in Eq. (\ref{start_zb_app}) as
\begin{eqnarray}\label{zb_smallw}
&&Z_B(y,y_0) = \frac{y e^{-\frac{y^2}{4}}}{4\sqrt{\pi}} \Bigg[ (y_0 \ln{y_0})  \\
&-& y_0  \left(\frac{{\cal Q}(y)}{y} -\gamma_E -\ln{(y)} - {\cal A}(y) \right) \Bigg] + {\cal O}(y_0^2 \ln y_0) \;,\nonumber
\end{eqnarray}
where ${\cal A}(y)$ is given by
\begin{equation}\label{expr_az}
{\cal A}(y) = \int_0^\infty du \frac{1}{(u-1)(u+1)^2} \left[\frac{1}{u} e^{-\frac{y^2}{4}\left(\frac{1}{u^2} - 1\right)} - u \right] \;.
\end{equation}

One can then obtain $Z(y,y_0) = Z^{(0)}(y,y_0) - \epsilon (Z_A(y,y_0) + Z_B(y,y_0))$ in the limit $y_0 \to 0$ from the Eqs. (\ref{Z0_smallw}), (\ref{za_small_w}) and (\ref{zb_smallw}) as:
\begin{equation}\label{structure_z}
Z(y,y_0) = y_0 (1 - \frac{\epsilon}{2}\ln{y_0}) \frac{y}{2 \sqrt{\pi}} e^{-\frac{y^2}{4}} -\epsilon y_0 \tilde Z_1(y) \;,
\end{equation}
where $\tilde Z_1(z)$ can be read off straightfowardly from Eqs.~(\ref{za_small_w}) and (\ref{zb_smallw}):
\begin{eqnarray}\label{expr_ztilde}
&&\tilde Z_1(y) = - 2\partial_y R^{(1)}(y)  \\
&&-\frac{\, y}{4 \sqrt{\pi}}  e^{-y^2/4}\left(\frac{{\cal Q}(y)}{y} - \gamma_E - \ln{y} - {\cal A}(y)\right) \nonumber \;.
\end{eqnarray}
This perturbative expansion (\ref{structure_z}) is fully consistent with the expected behavior~\cite{ZK95, zoia_semi_inf}
\begin{equation}
Z(y,y_0) \sim y_0^{\frac{\alpha}{2}} \tilde Z(y) \;, \; y_0 \to 0 \;, \tilde Z(y) = \tilde Z^{(0)}(y)  - \epsilon \tilde Z^{(1)}(y) \;,
\end{equation}
where $\tilde Z^{(1)}(y)$ is given in Eq. (\ref{expr_ztilde}). These integrals that enter the definition of $\tilde Z^{(1)}(y)$ can 
then be computed, for instance using Mathematica, to yield the following expression of $R_+(y)$ from Eq. (\ref{ratio_r+}): 
\begin{widetext}
\begin{eqnarray}\label{W_elementary}
&&R_+(y) = R_+^{(0)}(y)  \left(1 + \epsilon W_+(y) + {\cal O}(\epsilon^2)\right) \;,\\
&& W_+(y) = \frac{1}{48} \Bigg[60 -24 \gamma_E - 48 \log{2} - y^2(18-6 \gamma_E-12\log{2}) - 6(2-\sqrt{\pi}y)e^{\frac{y^2}{4}} \nonumber \\
&&+ y^2(y^2-4) \frac{y^4}{2} \,_2 F_2\left({1,1};{\frac{5}{2},3};\frac{y^2}{4}\right)  -\frac{24 \sqrt{\pi}}{y} e^{\frac{y^2}{4}} {\rm Erf}\left(\frac{y}{2} \right) - 3 \pi(y^2-2) {\rm Erfi}\left(\frac{y}{2}\right) + 3(y^2-4) {\rm Ei}\left(\frac{y^2}{4}\right) \Bigg] \;, \nonumber
\end{eqnarray}
\end{widetext}
where $R_+^{(0)}(y)$ is given in Eq. (\ref{Rp0}) and $\,_2 F_2\left({1,1};{\frac{5}{2},3};u\right)$ is a hypergeometric series~\cite{abramowitz}. From this expression (\ref{W_elementary}) it is straightforward to obtain the asymptotic behaviors given in the text (\ref{rplus_smallx}, \ref{rplus_largex}).

\end{document}